\documentclass[conf]{new-aiaa}
\usepackage[utf8]{inputenc}

\usepackage{graphicx}
\usepackage{amsmath}
\usepackage[version=4]{mhchem}
\usepackage{siunitx}
\usepackage{longtable,tabularx}
\usepackage{amsmath}
\usepackage{upgreek}
\usepackage{pdflscape}
\usepackage{listings}
\usepackage{multirow}
\usepackage[percent]{overpic}
\usepackage{multicol}
\usepackage{color}
\usepackage{stmaryrd}
\usepackage{xfrac}
\usepackage[capitalise]{cleveref}
\usepackage{booktabs}
\usepackage[section]{placeins} 
\usepackage[section]{algorithm}
\usepackage{calc}
\usepackage[titletoc]{appendix}
\usepackage{siunitx}
\usepackage{mathtools}
\usepackage{tabularx}
\usepackage{amsmath}
\usepackage{diagbox}

\usepackage{tikz-dimline}
\usepackage{graphicx}
\usepackage{graphics}
\usepackage{wrapfig}
\usepackage{float}
\usepackage{subfig}
\usepackage[percent]{overpic}
\usepackage{varwidth}
\usepackage{tikz}
\usepackage{pgfplots}
\usepackage{adjustbox}
\usetikzlibrary{arrows,matrix,positioning,fit}
\usetikzlibrary{shapes,positioning}
\usetikzlibrary{backgrounds}
\usepackage{tikz-layers}
\usepgfplotslibrary{fillbetween}
\usetikzlibrary{intersections}
\pgfplotsset{compat=1.14}
\usepgfplotslibrary{colorbrewer}
\usepgfplotslibrary{patchplots}
\usepgfplotslibrary[colorbrewer]
\usetikzlibrary{pgfplots.colorbrewer}
\usetikzlibrary[pgfplots.colorbrewer]
\usepgfplotslibrary{units}
\usetikzlibrary{spy}
\usepackage{pgfplotstable}
\usepackage{arrayjobx}
\graphicspath{ {./figs/} }
\usetikzlibrary{external}

\newlength\myheight
\newlength\mydepth
\settototalheight\myheight{Xygp}
\newcommand*\inlinegraphics[1]{%
  \settototalheight\myheight{Xygp}%
  \settodepth\mydepth{Xygp}%
  \raisebox{-\mydepth}{\includegraphics[height=\myheight]{#1}}%
}
\newcommand\orcid[1]{\href{https://orcid.org/#1}{\inlinegraphics{orcid_16x16.png}}}

\makeatletter
\def\BState{\State\hskip-\ALG@thistlm}
\makeatother

\title{Direct molecular gas dynamics simulations of re-entry vehicles via the Boltzmann equation}

\author{Tarik Dzanic\footnote{Postdoctoral fellow}}
\affil{Center for Applied Scientific Computing, Lawrence Livermore National Lab, Livermore, CA 94550, USA}
\author{Luigi Martinelli\footnote{Professor; Associate Fellow, AIAA}}
\affil{Department of Mechanical and Aerospace Engineering, Princeton University, Princeton, NJ 08544, USA}

\begin{document}

\maketitle

\begin{abstract}
This work explores the feasibility of performing three-dimensional molecular gas dynamics simulations of hypersonic flows such as re-entry vehicles through directly solving the six-dimensional nonlinear Boltzmann equation closed with the BGK (Bhatnagar--Gross--Krook) collision model. Through the combination of high-order unstructured spatial discretizations and conservative discrete velocity models as well as their efficient implementation on large-scale GPU computing architectures, we demonstrate the ability to simulate unsteady and non-equilibrium three-dimensional high-speed flows at a feasible computational cost through a unified numerical framework. We present the results of high-order simulations of the Apollo capsule at realistic re-entry conditions from the AS-202 mission flight path, including the steady non-equilibrium flow in the high-altitude regime at a Mach number of 22.7 and a Reynolds number of 43,000 as well as the unsteady turbulent flow in the low-altitude regime at a Mach number of 8 and a Reynolds number of 550,000. The results show the validity of the approach over the entire range of a typical re-entry trajectory from the rarefied to the continuum limit, the ability to directly resolve strong shocks profiles without numerical shock capturing techniques, and the ability of resolving small-scale unsteady flow structures in the inertial range.


\end{abstract}




\section{Nomenclature}

{\renewcommand\arraystretch{1.0}
\noindent\begin{longtable*}{@{}l @{\quad=\quad} l@{}}
$\alpha$ & angle of attack \\
$C$ & particle collision operator \\
$d$ & spatial dimensionality \\
$D$ & capsule diameter \\
$\Delta x$ & mesh spacing \\
$\delta$ & number of internal degrees of freedom\\
$E$ & macroscopic total energy \\
$f$ & particle distribution function \\
$\mathbf{f}$ & discrete particle distribution function \\
$F$ & reduced velocity particle distribution function \\
$F^I$ & common interface flux \\
$\mathbf{F}$ & flux vector \\
$g$ & equilibrium particle distribution function \\
$g_{\mathbf{u}}$ & reduced velocity equilibrium particle distribution function\\
$g_\zeta$ & reduced internal energy equilibrium particle distribution function\\
$\mathbf{g}$ & discrete equilibrium particle distribution function \\
$G$ & reduced internal energy particle distribution function \\
$\mathbf{h}$ & flux reconstruction correction function \\
$\mathbf{I}$ & identity matrix \\
$K$ & Kelvin \\
$Kn$ & Knudsen number \\
$Kn_{\text{GLL}}$ & gradient-length local Knudsen number \\
$\lambda$ & particle mean free path \\
$\Lambda$ & Lambda function \\
$m$ & velocity dimensionality \\
$M$ & Mach number \\
$\mu$ & dynamic viscosity \\
$N_e$ & number of mesh elements \\
$N_f$ & number of flux nodes \\
$N_s$ & number of solution nodes \\
$N_u$ & number of velocity nodes \\
$\Omega^\mathbf{u}$ & velocity domain \\
$\Omega^\mathbf{x}$ & spatial domain \\
$\Omega^{\mathbf{x}}_k$ & element in mesh \\
$P$ & macroscopic pressure \\
$Pr$ & Prandtl number \\
$\Pi$ & pressure tensor \\
$\mathbf{q}$ & macroscopic primitive variables \\
$\mathbf{Q}$ & macroscopic conserved variables \\
$\mathbf{Q}'$ & modified macroscopic conserved variables \\
$Re$ & Reynolds number \\
$\rho$ & macroscopic density \\
$\rho \mathbf{U}$ & macroscopic momentum \\
$t$ & physical time \\
$\tau$ & collision time scale \\
$\theta$ & macroscopic scaled temperature \\
$\theta_{\text{ref}}$ & reference macroscopic scaled temperature \\
$\theta_S$ & Sutherland scaled temperature \\
$\mathcal T$ & elliptic temperature tensor \\
$\mathbf{u}$ & velocity variable \\
$u_{\max}$ & maximum discrete particle velocity \\
$\mathbf{x}$ & spatial variable \\
$\mathbf{x}^s$ & spatial solution nodes \\
$\mathbf{x}^f$ & spatial flux nodes \\
$\zeta$ & internal energy variable \\

\end{longtable*}}
\section{Introduction}
\lettrine{T}{he} comprehensive investigation of aerothermodynamic behavior of spacecraft during atmospheric re-entry remains a critical component of the engineering development cycle, providing crucial information for structural and material design as well as guiding the development of thermal protection systems for the vehicles. However, accurate simulations of flows in high-altitude, hypersonic conditions can be an exceptionally difficult undertaking, primarily driven by the complexity of the multi-scale, non-equilibrium nature of the flow in these conditions. Moreover, prediction of laminar-turbulent transition in hypersonic boundary layers is paramount as it determines the heat transfer and friction drag, and the prediction of the dynamics of vehicles at hypersonic speeds may be substantially hampered by uncertainties in the modeling of turbulence and turbulent transition, which is still poorly understood~\cite{Schneider2006}. 

Standard computational fluid dynamics approaches, which rely on solving the Navier--Stokes equations governing the dynamics of fluid flows, inherently assume that the fluid in question can be treated as continuum, such that the flow can be adequately described through its macroscopic state (e.g., density, momentum, energy). Under re-entry conditions, this assumption can become severely ill-posed as the flow experiences strong thermodynamic non-equilibrium, and its behavior can drastically vary from the predictions given by governing equations that do rely on the continuum assumption. In these cases, to provide an accurate description of the underlying flow physics, it is necessary to revert to more general governing equations that are derived from the kinetic theory of gases which can account for the non-equilibrium nature of the flow. 

The governing equations of molecular gas dynamics, which underpin the macroscopic behavior of the flow, provide a means of accurately describing non-equilibrium flows while seamlessly recovering the hydrodynamic equations in the continuum limit. As such, these models are the only ones truly capable of describing the flow physics across the entire flight path, from the high-altitude, rarefied regime to the low-altitude, continuum regime. One such approach is provided by approximation of the Boltzmann equation, which gives a statistical description of particle transport and collision at the mesoscale level. However, the complexity of the Boltzmann equation, primarily driven by the computational cost associated with its high-dimensional nature and the difficulty of approximating the collision process, severely limits its applicability. While more complex flow problems have been attempted with probabilistic methods such as direct simulation Monte Carlo approaches, the direct deterministic approximation of the Boltzmann equation has been widely considered computationally intractable for practical high-speed applications, and their use has generally been restricted to simpler, two-dimensional flows.

The aim of this work is to demonstrate the feasibility of performing three-dimensional molecular gas dynamics simulations of high-speed flows such as re-entry vehicles through directly solving the Boltzmann equation. This capability stems from recent developments by the authors for approximating the Boltzmann equation closed with the Bhatnagar--Gross--Krook (BGK) model which combines highly-efficient, high-order unstructured discontinuous spectral element method with a conservative discrete velocity model, the combination of which results in a low-memory, compute-heavy numerical algorithm which is well-suited for large-scale GPU computing architectures. As such, the authors have presented results of complex three-dimensional turbulent flows and high-speed flows simulated without any numerical shock capturing approaches by directly approximating the Boltzmann equation, both of which are entirely feasible on modern computing hardware~\citep{Dzanic2023,Dzanic2023b}. These results pave the way for investigating non-equilibrium flow around spacecraft in re-entry conditions through deterministic approximations of molecular gas dynamics.


In this work, we present the results of high-order three-dimensional numerical experiments of the flow around an AS-202 Apollo capsule at realistic re-entry conditions obtained through direct approximation of the Boltzmann--BGK equation. The first simulation is performed in the high-altitude region at a Mach number of 22.7 and a Reynolds number of 43,000, which results in strong non-equilibrium flow that is not accurately resolved by continuum fluid dynamics approximations~\citep{Wright2006}. The second simulation is performed in the low-altitude region at a Mach number of 8 and a Reynolds number of 550,000, which results in complex unsteady flow separation and turbulent wake flow. While three-dimensional re-entry flows have been attempted using kinetic/hybrid schemes in works such as that of \citet{Yuan2020}, \citet{Zhang2023}, and \citet{Chen2020}, these approaches have only been used for steady flows and with low-order numerical schemes. To the authors' knowledge, this work is the first to showcase the ability to directly solve the Boltzmann equation for unsteady hypersonic turbulent flows, achieving so with a high-order unstructured numerical scheme. The results presented in this work, in conjunction with previous work by the authors, highlight some unique properties of the proposed model, such as: i) validity over the entire range of a typical re-entry trajectory, from the rarefied to the continuum limit; ii) ability of directly resolving strong shocks profiles without numerical shock capturing techniques; and iii) ability of resolving small-scale unsteady flow structures in the inertial range. The overarching goal of this work is to showcase a novel approach for tackling complex high-speed flow problems which is more suited for handling the high-temperature aerothermodynamic effects that are encountered in these applications and may lead to enhancing our fundamental understanding of flow phenomena such as transition to turbulence through a radically different perspective of flow dynamics encoded in the high-dimensional solution of the Boltzmann equation. 

\section{Methodology}\label{sec:methodology}

\subsection{Governing equations}
The Boltzmann equation, which gives a deterministic description of molecular gas dynamics, can be given as
\begin{equation}\label{eq:boltzmann}
    \partial_t f (\mathbf{x}, \mathbf{u}, \zeta, t) + \mathbf{u} {\cdot} \nabla f = \mathcal C(f, f'),
\end{equation}
where $\mathbf{x}$ represents a $d$-dimensional physical space, $\mathbf{u}$ represents an $m$-dimensional physical space, $\zeta$ represents a scalar internal energy, $f (\mathbf{x}, \mathbf{u}, \zeta, t)$ is a scalar particle distribution function, and $\mathcal C(f, f')$ is some collision operator that accounts for intermolecular interactions~\citep{Cercignani1988}. The distribution functions represents a probability measure for a particle existing at a given location $\mathbf{x}$ traveling at a given velocity $\mathbf{u}$ with an internal energy $\zeta$. The moments of this distribution function recover the macroscopic state of the system, i.e., \begin{equation}\label{eq:moments}
    \mathbf{Q}(\mathbf{x}, t) = \left[\rho, \rho \mathbf{U}, E \right]^T = 
    \int_{\mathbb R^m} \int_{0}^{\infty} f (\mathbf{x}, \mathbf{u}, \zeta, t)\ \boldsymbol{\psi} (\mathbf{u}, \zeta) \ \mathrm{d}\zeta\ \mathrm{d}\mathbf{u} 
\end{equation}
where $\rho$ is the density, $\rho \mathbf{U}$ is the momentum vector, $E$ is the total energy, and $\boldsymbol{\psi} (\mathbf{u}, \zeta) \coloneqq [1, \mathbf{u}, (\mathbf{u}\cdot\mathbf{u})/2 + \zeta]^T$ is the vector of collision invariants. From this, a vector of primitive variables can be defined as $\mathbf{q} = [\rho, \mathbf{U}, P]$, where $\mathbf{U} = \rho \mathbf{U}/\rho$ is the macroscopic velocity, $P = (\gamma - 1)(E - \rho\mathbf{U}\cdot\mathbf{U}/2)$ is the pressure, and $\gamma$ is the specific heat ratio. We use the notation $\mathbf{u}$ to denote the microscopic velocities and $\mathbf{U}$ to denote the macroscopic velocities. 

The collision operator is approximated using the Bhatnagar--Gross--Krook (BGK) model~\citep{Bhatnagar1954}, given as 
\begin{equation}
    \mathcal C (f, f') \approx \frac{g(\mathbf{x}, \mathbf{u}, \zeta, t) - f(\mathbf{x}, \mathbf{u}, \zeta, t)}{\tau},
\end{equation}
where $g(\mathbf{x}, \mathbf{u}, \zeta, t)$ is the equilibrium distribution function (EDF) and $\tau$ is the collision time scale. To recover a non-unit Prandtl number and account for the effects of internal degrees of freedom, the EDF is computed through the ellipsoidal BGK (ES-BGK) approach~\citep{Holway1966} in conjunction with the internal energy model of \citet{Baranger2020}. The EDF can be represented as a product of the monatomic ES-BGK EDF $g_\mathbf{u}$ and the internal energy EDF $g_\zeta$ as 
\begin{equation}
g(\mathbf{x}, \mathbf{u}, \zeta, t) = g_\mathbf{u}(\mathbf{x}, \mathbf{u}, t) \times g_\zeta (\mathbf{x}, \zeta, t).
\end{equation}
The ES-BGK EDF can be computed as
\begin{equation}
    g_\mathbf{u} (\mathbf{x}, \mathbf{u}, t) = 
    \frac{\rho (\mathbf{x}, t)}{\sqrt{\text{det}(2 \pi \mathcal{T})}}\exp \left [-\frac{1}{2}\left[\mathbf{u} - \mathbf{U}(\mathbf x, t) \right]^T \mathcal{T}^{-1} \left[\mathbf{u} - \mathbf{U}(\mathbf x, t)\right]\right],
\end{equation}
where the temperature tensor $\mathcal T$ is computed as a combination of the temperature $\theta = P/\rho$ and the density-normalized pressure tensor $\Pi$, i.e.,
\begin{equation}
    \mathcal{T} = \frac{1}{Pr} \theta \mathbf{I} + (1 - \frac{1}{Pr}) \frac{\Pi}{\rho}, \quad \text{where} \quad 
    \Pi_{ij} = \int_{\mathbb R^m} f (u_i - U_i)(u_j - U_j)\ \mathrm{d}\mathbf{u}.
\end{equation}
The Prandtl number is set as $Pr = 0.71$. From this model, the collision time scale can be related to the dynamic viscosity $\mu$ as
\begin{equation}
    \tau = \frac{1}{Pr}\frac{\mu}{P}.
\end{equation}
In this work, the Sutherland viscosity model~\citep{Sutherland1893} is used, given by the relation
\begin{equation}
    \mu = \mu_{\text{ref}}\left(\frac{\theta}{\theta_{\text{ref}}} \right)^{\frac{3}{2}}\frac{\theta_{\text{ref}} + \theta_S}{\theta + \theta_S},
\end{equation}
where $(\cdot)_\text{ref}$ denotes the reference quantities and $\theta_S$ denotes the (scaled) Sutherland temperature. 

The internal energy EDF is computed as
\begin{equation}
    g_\zeta (\mathbf{x}, \zeta, t) = \Lambda(\delta) \left (\frac{\zeta}{\theta(\mathbf{x}, t)} \right)^{\frac{\delta}{2} - 1} \frac{1}{\theta(\mathbf{x}, t)}\exp \left(-\frac{\zeta}{\theta(\mathbf{x}, t)} \right) ,
\end{equation}
where $\delta \geq 0$ is the number of internal degrees of freedom and $\Lambda (\delta) = 1/\Gamma(\delta/2)$ is a normalization factor to ensure that $g$ recovers the same macroscopic state as $f$. The number of internal degrees of freedom are not necessarily fixed nor required to be an integer, such that spatially varying and temperature-dependent internal degrees of freedom are possible. However, the addition of an internal energy model further increases the dimensionality of the Boltzmann equation, which would significantly increase the associated computational cost. Since it is typically not necessary to know the actual distribution within the internal energy domain, only its effect on the macroscopic quantities, it is beneficial to utilize a reduced distribution technique for the internal energy domain~\citep{Baranger2020}. In this approach, the distribution function $f(\mathbf{x}, \mathbf{u}, \zeta, t)$ (and its evolution) is integrated with respect to the internal energy variable $\zeta$ to yield a pair of distribution functions $[F,G]^T$ as
\begin{equation}
    \begin{bmatrix}
        F(\mathbf{x}, \mathbf{u}, t) \\
        G(\mathbf{x}, \mathbf{u}, t)
    \end{bmatrix}
    =
    \int_0^\infty
    \begin{bmatrix}
        1 \\
        \zeta
    \end{bmatrix}
    f(\mathbf{x}, \mathbf{u}, \zeta, t)\ \mathrm{d}\zeta.
\end{equation}
Critically, these reduced distribution functions are only defined over the spatial and velocity domains, not the internal energy domain, such that the computational cost and memory requirements are only twice the cost of the monatomic case. The evolution of these reduced distribution function can be computed as 
\begin{equation}
    \partial_t 
    \begin{bmatrix}
        F(\mathbf{x}, \mathbf{u}, t) \\
        G(\mathbf{x}, \mathbf{u}, t)
    \end{bmatrix} 
    + \mathbf{u}\cdot\nabla
    \begin{bmatrix}
        F(\mathbf{x}, \mathbf{u}, t) \\
        G(\mathbf{x}, \mathbf{u}, t)
    \end{bmatrix} 
    = \frac{1}{\tau}
    \begin{bmatrix}
        \hphantom{\frac{\delta \theta}{2}}g_\mathbf{u} (\mathbf{x}, \mathbf{u}, t) - F(\mathbf{x}, \mathbf{u}, t)  \\
        \frac{\delta \theta}{2} g_\mathbf{u} (\mathbf{x}, \mathbf{u}, t) - G(\mathbf{x}, \mathbf{u}, t)
    \end{bmatrix},
\end{equation}
where it can be seen that the equilibrium distribution function of the two components varies only by a factor of $\frac{\delta \theta}{2}$. From this formulation, the macroscopic quantities can be computed as
\begin{equation}
    \begin{bmatrix}
        \rho \\
        \rho \mathbf{U}\\
        E
    \end{bmatrix}
    =
    \int_{\mathbb R^m}
    \begin{bmatrix}
        1 \\
        \mathbf{u}\cdot\mathbf{u} \\
        (\mathbf{u}\cdot\mathbf{u})/2 
    \end{bmatrix}
    F
    +
    \begin{bmatrix}
        0 \\
        0 \\
        1 
    \end{bmatrix}
    G\
    \mathrm{d}\mathbf{u}.
\end{equation}
Similarly, the pressure tensor $\Pi$ can be computed as
\begin{equation}
    \Pi_{ij} = \int_{\mathbb R^m} F (u_i - U_i)(u_j - U_j)\ \mathrm{d}\mathbf{u}.
\end{equation}

\subsection{Numerical method}
\begin{figure}[tbhp]
    \begin{centering}
    \adjustbox{width=0.6\linewidth, valign=b}{\input{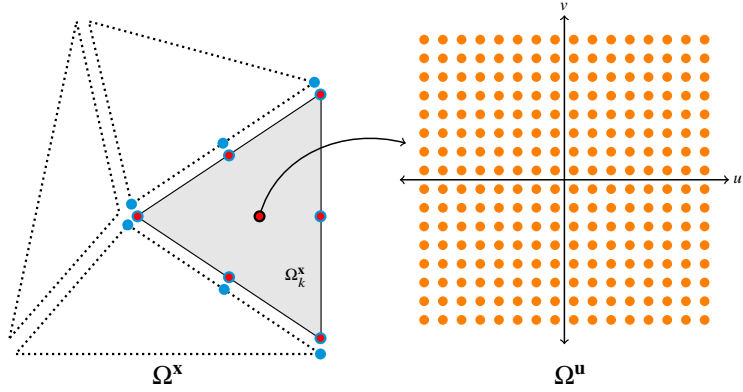}}
    \caption{\label{fig:scheme} Schematic of a two-dimensional phase space discretization using an unstructured spatial domain $\Omega^{\mathbf{x}}$ with $\mathbb P_2$ elements and a velocity domain $\Omega^{\mathbf{u}}$ with $N_u = 16^2$. Circles denote the spatial solution nodes (red), interface flux nodes (blue), and velocity space nodes (orange), respectively.}
    \end{centering}
    \end{figure}

The underlying numerical scheme in this work is the method of \citet{Dzanic2023}. This approach combines a high-order spatial discretization with a discretely-conservative velocity model, the combination of which results in a low-memory, compute-heavy numerical algorithm which is well-suited for large-scale GPU computing architectures. A brief overview of this scheme is presented here, with a schematic of the approach shown in \cref{fig:scheme}.

The spatial domain is discretized using a high-order flux reconstruction scheme~\citep{Huynh2007}, a generalization of the nodal discontinuous Galerkin method~\citep{Hesthaven2008DG}. When combined with a nodal velocity discretization, the coupled spatial and velocity domains reduce to a set of decoupled advection equations for each nodal velocity point $\mathbf{u}_0$ with a nonlinear source term $S$, i.e.,
\begin{equation}\label{eq:transport}
    \partial_t f (\mathbf{x}, \zeta, t) +  \boldsymbol{\nabla}\mathbf{F}(f)   f = S, \quad \mathbf{F}(f) = \mathbf{u}_0 f.
\end{equation}
For this scheme, the spatial domain $\Omega^{\mathbf{x}}$ is partitioned into $N_e$ elements $\Omega^{\mathbf{x}}_k $ such that $\Omega^{\mathbf{x}} = \bigcup_{N_e}\Omega^{\mathbf{x}}_k$ and $\Omega^{\mathbf{x}}_i\cap\Omega^{\mathbf{x}}_j=\emptyset$ for $i\neq j$, shown on the left-hand side of \cref{fig:scheme}. The distribution function within each element is approximated by an order $\mathbb P_p$ polynomial as
\begin{equation}
    f (\mathbf{x}) = \sum_{i = 1}^{N_s} f (\mathbf{x}^s_i) {\phi}_i (\mathbf{x}),
\end{equation}
where $\mathbf{x}^s_i \in \Omega^{\mathbf{x}}_k \ \forall \ i \in \{1,..., N_s\}$ is a set of $N_s$ solution nodes and ${\phi}_i (\mathbf{x})$ is a set of nodal basis functions. The flux is then computed as 
\begin{equation}
    \mathbf{F}(\mathbf{x}) = \mathbf{u}_0 f (\mathbf{x}) + \sum_{i = 1}^{N_f} \left[F^I_i - f (\mathbf{x}^f_i) \mathbf{u}_0\cdot \mathbf{n}_i \right] \mathbf{h}_i (\mathbf{x}).
\end{equation}
Here, $\mathbf{x}^f_i \in \partial \Omega^{\mathbf{x}}_k \ \forall \ i \in \{1,..., N_f\}$ is a set of $N_f$ interface flux nodes, $\mathbf{n}_i$ is their associated outward-facing normal vector, $F^I_i$ is the upwind numerical flux at the interface, and $\mathbf{h}_i$ is the correction function used to recover the nodal discontinuous Galerkin approach~\citep{Huynh2007, Hesthaven2008DG}. The upwind numerical flux is simply computed as
\begin{equation}
    F_i^I = \begin{cases}
    u_n f_i^-, \quad \quad \mathrm{if} \ u_n > 0,\\
    u_n f_i^+, \quad \quad \mathrm{else},
    \end{cases}
\end{equation}
where $u_n = \mathbf{u}_0 \cdot \mathbf{n}_i$ and the superscripts $-$ and $+$ denote the interior value (from the element of interest) and the exterior value (from the face-adjacent element) of the solution at the interface, respectively.

The velocity domain is approximated by a set of discrete velocity nodes distributed across a Cartesian grid in velocity space, shown on the right-hand side of \cref{fig:scheme}. The extent of the velocity domain is chosen such as to adequately encompass the majority of the distribution function, typically taken as some factor of the maximal thermal velocity and macroscopic velocity. The velocity domain is then taken as $[-u_{\max}, u_{\max}]^m$, where $u_{\max}$ is the maximum velocity extent. The governing equations are then solved for across the discrete velocity nodes, each corresponding to a spatially decoupled advection equation as in \cref{eq:transport}. These spatially decoupled systems are linked together by the collision operator which acts as a nonlinear source term across the velocity space. 

However, the BGK (and ES-BGK) collision operators require integrating the particle distribution function across the velocity space to compute the moments. If this integration is not exact for the moments corresponding to the macroscopic state, the equilibrium distribution function $g$ may not possess the same macroscopic state as $f$. As a result, the scheme would not be conservative with respect to the macroscopic state, and these conservation errors can ultimately result in numerical instabilities~\citep{Dzanic2023}. A typical remedy for this problem is to increase the resolution in velocity space to reduce the integration errors or choose the velocity basis and/or nodal quadrature points judiciously such as to integrate particular moments of the distribution function exactly (e.g., moment-based approximations/Gauss--Hermite quadrature). However, this either results in a prohibitively large computational cost or a velocity discretization which may be ill-suited, particularly so for high Mach number flows~\citep{Cai2019,Heyningen2021}. 

Instead, we utilize the discrete velocity model of \citet{Mieussens2000} which ensures that the scheme remains discretely conservative regardless of the resolution and choice of velocity nodal distribution, which can drastically reducing the computational cost and resolution requirements of standard nodal velocity approximations and offer a more general and robust approach than moment/Gauss--Hermite based methods. In brief, this approach requires iteratively solving for a modified equilibrium distribution function that possesses the same macroscopic moments as $f$ under discrete integration. As a result, this can transform a memory-heavy velocity space approximation into a lower memory, compute-heavy task which is optimal for modern GPU architectures. 

The goal of the discrete velocity model is to find this modified EDF $\mathbf{g}'$ which satisfies the \emph{discrete compatibility condition}, i.e.,
\begin{equation}\label{eq:discrete_compatibility}
    \mathbf{M} \cdot \left[\mathbf{f} \otimes \boldsymbol{\psi} \right] = \mathbf{M} \cdot \left[\mathbf{g}' \otimes \boldsymbol{\psi} \right].
\end{equation}
where 
\begin{equation}
    \mathbf{M} \cdot \mathbf{f} \approx \int_{\mathbb R^m} f (\mathbf{u}) \ \mathrm{d}\mathbf{u}
\end{equation}
is a discrete integration operator with strictly-positive entries. 
Due to the use of a uniform Cartesian distribution in the velocity domain, we utilize the trapezoidal rule which offers spectral convergence for smooth, compactly-supported, periodic functions on uniform grids~\citep{Trefethen2014}. It was shown in \citet{Mieussens2000} that the EDF which satisfies both the discrete compatibility condition and the discrete form of the H-theorem is equivalent to the analytic EDF formed around a modified macroscopic state $\mathbf{Q}'$ (i.e., $\mathbf{g}(\mathbf{Q}')$) which converges to the true macroscopic state in the limit of infinite resolution. As a result, finding this modified EDF typically can be framed simply as the solution of a $d+2$ dimensional nonlinear system, irrespective of the dimensionality of the velocity space -- note that in some scenarios, the ES-BGK model may require a nonlinear solve of a marginally higher dimensionality if the velocity space does not possess certain symmetry properties~\citep{Mieussens2000JCP}. This low-dimensional, compute-heavy optimization process can be rapidly performed on GPUs using root-finding approaches such as Newton's method.

\subsection{Geometry \& flow conditions}

\begin{figure}[htbp!]
    \centering
    \adjustbox{width=0.4\linewidth, valign=b}{\input{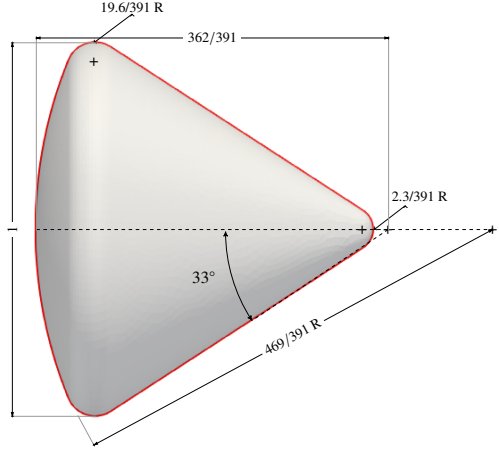}}
    \caption{Cross-section schematic of the AS-202 \textit{Apollo} capsule geometry normalized to unit diameter.}
    \label{fig:geo}
\end{figure}

The numerical experiments focus on the three-dimensional simulation of the AS-202 \textit{Apollo} capsule at realistic operating conditions. A schematic of the cross-sectional profile of the capsule geometry is shown in \cref{fig:geo}, normalized to a unit non-dimensional diameter corresponding to a physical diameter of $3.912$ m. The capsule geometry consists of a blunt spherical forebody followed by a $33^{\circ}$ conical afterbody. We use the notation that fore/aft is with respect to the flow-oriented reference frame, not with respect to the reference frame of the astronauts which are positioned in the opposite direction. 

The operating conditions are taken from the AS-202 experimental flight data of \citet{Hillje1967} as reported by \citet{Wright2006}. The flight data is reported over a range of flight times corresponding to $4350-5200$ seconds after launch, during which the capsule travels from an altitude of $122$ to $8$ kilometers. During this time period, the operating Mach number steadily decreases from $M = 31$, while the Reynolds number (based on the capsule diameter) increases to a maximum of $Re = 10^7$. The first simulation operating conditions were chosen to match the experimental flight data at 4700 seconds, which corresponds to an altitude of $77.2$ kilometers, a Mach number of $M = 22.7$, a Reynolds number of $Re = 43,000$, a freestream temperature of $203$ K, an angle of attack of $\alpha = 18.5^{\circ}$, and a zero side-slip angle. These conditions were chosen as the high Mach number and moderate Reynolds number result in strong non-equilibrium flow behavior, such that continuum fluid dynamics approximations yield erroneous predictions~\citep{Wright2006}. The second simulation operating conditions were chosen similarly to the experimental flight data at 5000 seconds, which corresponds to an altitude of $44.8$ kilometers and a Mach number of $M = 8$. The Reynolds number was set to $Re = 550,000$ to yield unsteady turbulent flow in the continuum regime, although this was marginally less than the true experimental Reynolds number which was on the order of $Re = 10^6$. As the purpose of this second experiment was primarily to show the validity of the approach for unsteady flows across the continuum regime instead of exactly matching the experimental conditions, the freestream temperature, angle of attack, and side-slip angle were kept the same as with the first simulation. 

We note here that the numerical experiments in this work are primarily meant to represent a preliminary proof of capability of directly solving the Boltzmann equation for such flows. As such, certain aspects of the flow as neglected, although these are strictly modeling assumptions and do not impact the computational feasibility of the proposed approach. In particular, we neglect the effects of multi-species reactions/chemical non-equilibrium and vibrational degrees of freedom, both of which were considered in the numerical experiments of \citet{Wright2006}, as well as ionization, shock-layer radiation, and material ablation. Additionally, the wall is simply assumed to be isothermal with a temperature equal to the freestream, whereas the work of \citet{Wright2006} assumes radiative equilibrium with surface catalysis effects which is more representative of the experimental flow conditions. As such, the results in this work are primarily used for qualitative comparison of the flow fields to the work of \citet{Wright2006}. A direct comparison of more complex flow effects such as near-wall flow behavior and an evaluation of the accuracy of predicting quantitative metrics such as wall heat transfer effects are contingent on future work in employing these additional modeling assumptions in the proposed approach.

\subsection{Implementation \& problem setup}
\begin{figure}[htb!]
        \centering
        \adjustbox{width=0.7\linewidth,valign=b}{\includegraphics{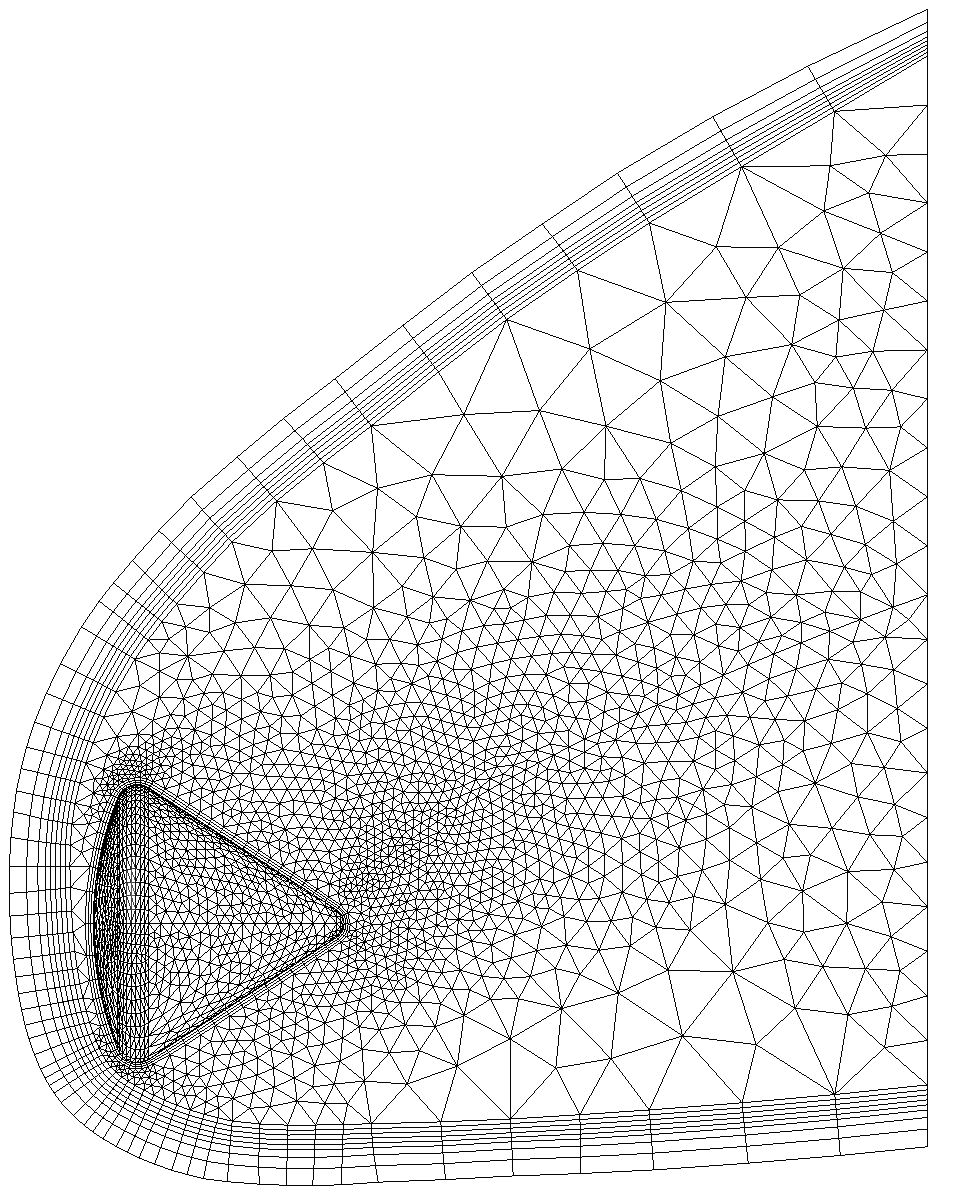}}
        \caption{\label{fig:mesh} Visualization of the high-order finite element mesh on capsule surface and symmetry plane.} 
    \end{figure}

Due to the zero side-slip angle, a half-domain was considered with symmetry enforced along the pitch plane ($z = 0$). An initial highly-resolved unstructured mesh was generated upon which the Navier--Stokes equations were solved to give an initial indication on the flow field in the problem. A coarser shock-fitted mesh was then generated for the Boltzmann--BGK solver, the results of which were used to guide further mesh improvements. After several iterations of solving the flow field via the Boltzmann--BGK approach and adapting the preliminary meshes, the mesh shown in \cref{fig:mesh} was generated and used for the resulting numerical experiments. We note here that the mesh was fitted primarily for the high-altitude/Mach simulation, and the same mesh was used for both simulations.

The mesh consisted of a unstructured triangular surface region with a near-wall prismatic boundary layer region. The domain was primarily filled with tetrahedral elements except near the shock region where shock-aligned prismatic elements were used. The mesh consisted of $14,033$ primastic elements and $25,680$ tetrahedral elements with a minimum mesh spacing of $\Delta x = 4\cdot{10^{-3}}$. This resolution level was chosen to mimic the mesh used in \citet{Wright2006}, with approximately 1.3 million spatial degrees of freedom at the operating order for the high-altitude case and 2.4 million spatial degrees of freedom for the low-altitude case, which, when combined with the high-order spatial accuracy of the proposed approach, is expected to yield more performant resolving power. Note that the mesh appears significantly coarser than a finite volume mesh of equivalent resolution due to the multiple degrees of freedom per element in high-order finite element methods (e.g., a single prismatic element using a third-order polynomial approximation contains 40 degrees of freedom), such that the effective resolution is on the order of $\Delta x/(p+1)$. To account for the coarseness of the mesh in high-order finite element methods, we utilize quadratic elements in the near-wall region to better account for the curvature of the geometry. 

The operating approximation order was set to $\mathbb P_3$ for the high-altitude case and $\mathbb P_4$ for the low-altitude case, yielding a spatially fourth-order and fifth-order accurate solution, respectively, and the solution/flux points were placed at the $\alpha$-optimized quadrature points~\citep{Hesthaven2008DG} for simplex elements and Gauss--Lobatto quadrature points for tensor-product elements. For the high-altitude/Mach simulation, the velocity space resolution was set to $N_u = 24^3$, resulting in a total of 30 billion degrees of freedom. The velocity domain extent was set to $\Omega^{\mathbf{u}} = [-0.6, 1.4] \times [-0.8, 1.2] \times [-1, 1]$. We note here that results obtained with lower velocity space resolution $N_u = 20^3$ showed marked similarities in the macroscopic flow fields. Therefore, it may be surmised that the flow was effectively converged in the velocity domain due to the spectral convergence properties of the approximation, although quantitative convergence analysis was not performed. For the low-altitude/Mach simulation, the velocity space resolution was set to $N_u = 16^3$ due to the lower resolution requirements at lower Mach numbers, and the velocity domain extent was set identically to the high-altitude simulation. Due to the reduced velocity space resolution in the low-altitude case, the operating order was increased from $\mathbb P_3$ to $\mathbb P_4$, such that the total degrees of freedom were approximately the same.

The initial conditions of the flow were normalized such as to yield a unit density and velocity, 
    \begin{subequations}
        \begin{align}
            \rho &= 1,\\
            u &= \cos(\alpha), \\
            v &= \sin(\alpha), \\
            w &= 0,\\
            P &= P_{\text{ref}},
        \end{align}
    \end{subequations}
with the pressure and reference viscosity set such as to yield the appropriate freestream Mach and Reynolds numbers, i.e.,
    \begin{subequations}
        \begin{align}
             P_{\text{ref}} = \frac{1}{\gamma M^2},\\
             \mu_{\text{ref}} = \frac{1}{Re}.
        \end{align}
    \end{subequations}
The reference scaled temperature was then set to $\theta_{\text{ref}} = P_{\text{ref}}/\rho_{\text{ref}} = P_{\text{ref}}$ and the Sutherland scaled temperature was set to $\theta_S = \theta_{\text{ref}} \frac{110.4}{203}$ which corresponds to a physical Sutherland temperature of $110.4$ K and a physical freestream temperature of $203$ K. Diffuse wall boundary conditions were enforced on the capsule surface, with the wall temperature set equal to the freestream/reference temperature. At the symmetry plane, symmetry was enforced using specular (reflecting) wall boundary conditions. At the farfield, Dirichlet boundary conditions were used with the boundary state set equal to the initial conditions. All boundary conditions were enforced weakly through the standard approach in discontinuous spectral element methods where an exterior state is given and common interface flux is computed via the Riemann solver. As such, upwinding was performed naturally at the particle level.

To reduce the unnecessary computational cost in resolving the initial transients in the solution, a Navier--Stokes flow field was first computed on the mesh and advanced until a quasi-steady state was achieved using the HLL Riemann solver~\citep{Harten1983} and the entropy filtering approach~\citep{Dzanic2022} for shock capturing. The Boltzmann solution was then initialized from this flow field by setting the initial distribution function state to the EDF corresponding to the macroscopic state of the Navier--Stokes flow field. The operating order was then gradually increased and the flow as computed by the Boltzmann--BGK approach was allowed to develop until the operating order was reached. For the high-altitude case, the flow was then advanced to a steady state, after which post-processing of the results was performed. For the low-altitude case, the primary purpose was to showcase the ability to compute unsteady flows. As such, we simply present results in terms of an instantaneous flow snapshot after the initial flow transients have been convected away.

Computations were performed using the PyFR codebase~\citep{Witherden2014} as the underlying solver. For the given problem and number of degrees of freedom, the memory requirements could be achieved with 16 NVIDIA H100 GPUs. The simulations in this work were however performed with 80 NVIDIA V100 GPUs. The total compute time for the high-altitude steady case was approximately 2 days, corresponding to roughly 3800 GPU hours. We note here that a sizeable portion of this compute time is spent advancing the solution to a steady state using an explicit time integration method. As such, the approach is better suited towards computing unsteady flow features~\citep{Dzanic2023b}, and for the low-altitude unsteady case, the compute time was approximately 11 hours (900 GPU hours) per characteristic flow time which corresponds to one flow over diameter. 

\section{Results}\label{sec:results}
\subsection{High-altitude conditions}
\subsubsection{Macroscopic flow cross-sections}
\begin{figure}[htbp!]
    \centering
    \subfloat[Density]{\includegraphics[width=0.48\textwidth]{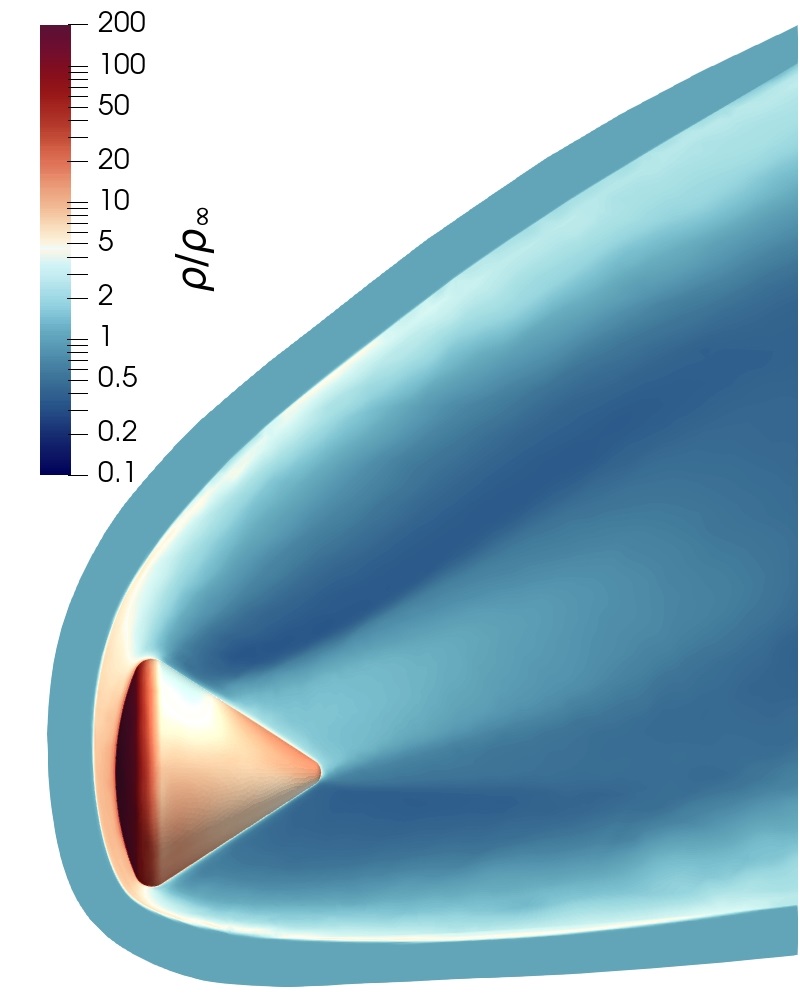}}
    \subfloat[Pressure]{\includegraphics[width=0.48\textwidth]{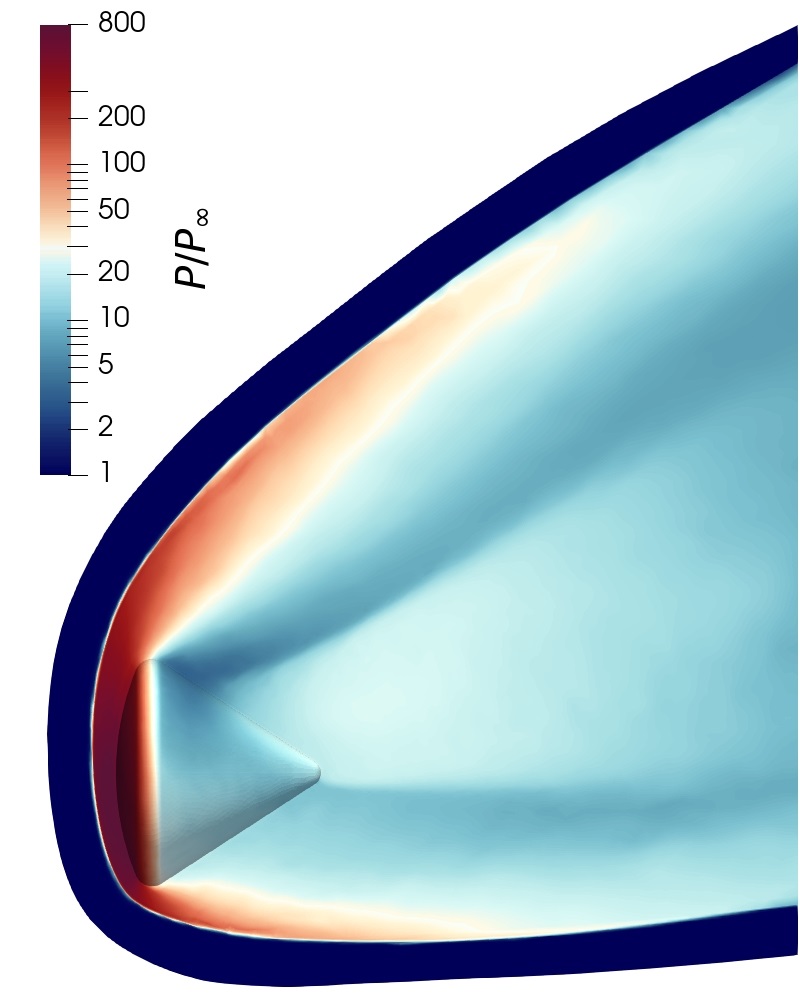}}
    \newline
    \subfloat[Velocity]{\includegraphics[width=0.48\textwidth]{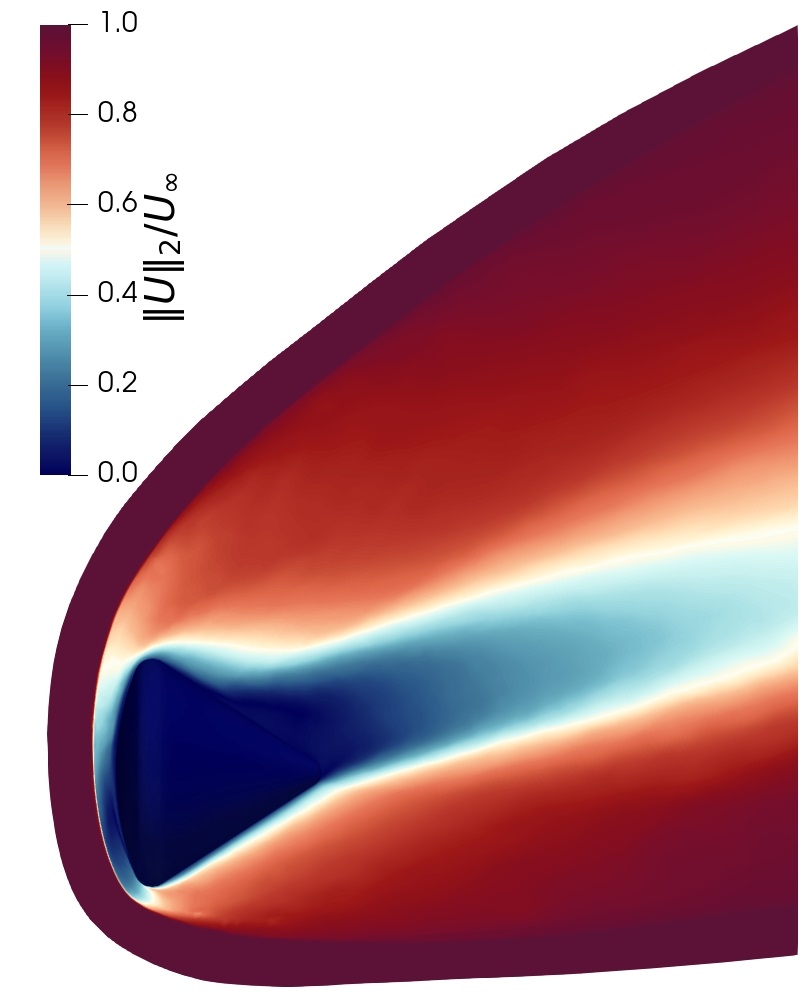}}
    \subfloat[Temperature]{\includegraphics[width=0.48\textwidth]{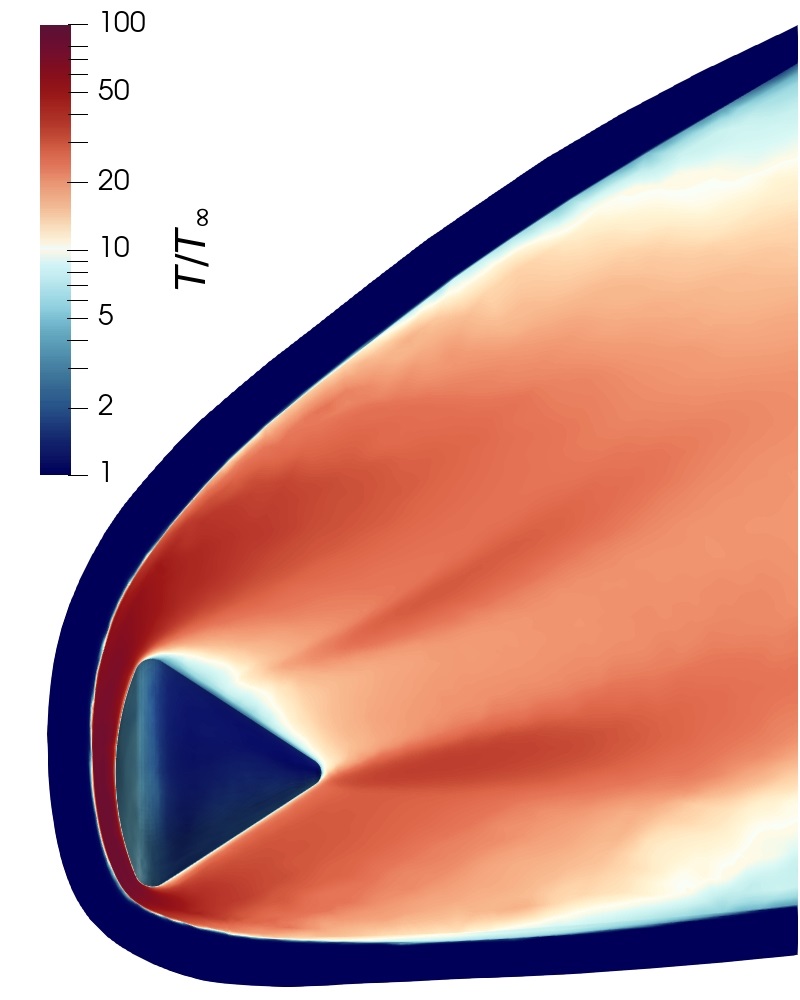}}
    \newline
    \caption{Contours of macroscopic normalized density (top left), pressure (top right), velocity magnitude (bottom left), and temperature (bottom right) on the pitch plane ($z = 0$) and the capsule surface for the high-altitude flow conditions. }
    \label{fig:sol}
\end{figure}

The results for the steady, high-altitude, high Mach case were first analyzed with respect to the macroscopic flow state on the symmetry plane cross-section ($z = 0$), with the contours of density, pressure, velocity magnitude, and temperature shown in \cref{fig:sol} normalized by the freestream values. The quintessential bow shock was evident in the flow field with large jumps in the density, pressure, and temperature across the shock. Due to the ability of the Boltzmann equation to directly resolve shock structures without any \text{ad hoc} numerical stabilization approach, the solution in the vicinity of the shock was well-behaved even with the use of a high-order numerical scheme in the presence of strong shocks. Density ratios of $\mathcal O (10)$ and pressure jumps of $\mathcal O (10^2)$ were observed across the shock, which align with the predictions of the normal shock equations. The highest density and pressure ratios were observed at the capsule surface on the windward side, with density and pressure ratios of over 200 and 800, respectively. 

A strong separation region was observed aft of the capsule surface on the leeward side, with low velocity magnitude and high density and pressure. This separation region largely corresponded to the location of the recompression shocks in the wake, with strong density and pressure gradients aligning with the edge of the separation region. These strong gradients were also well-resolved by the numerical scheme. 

\subsubsection{Surface flow behavior}
\begin{figure}[htbp!]
    \centering
    \includegraphics[width=0.6\textwidth]{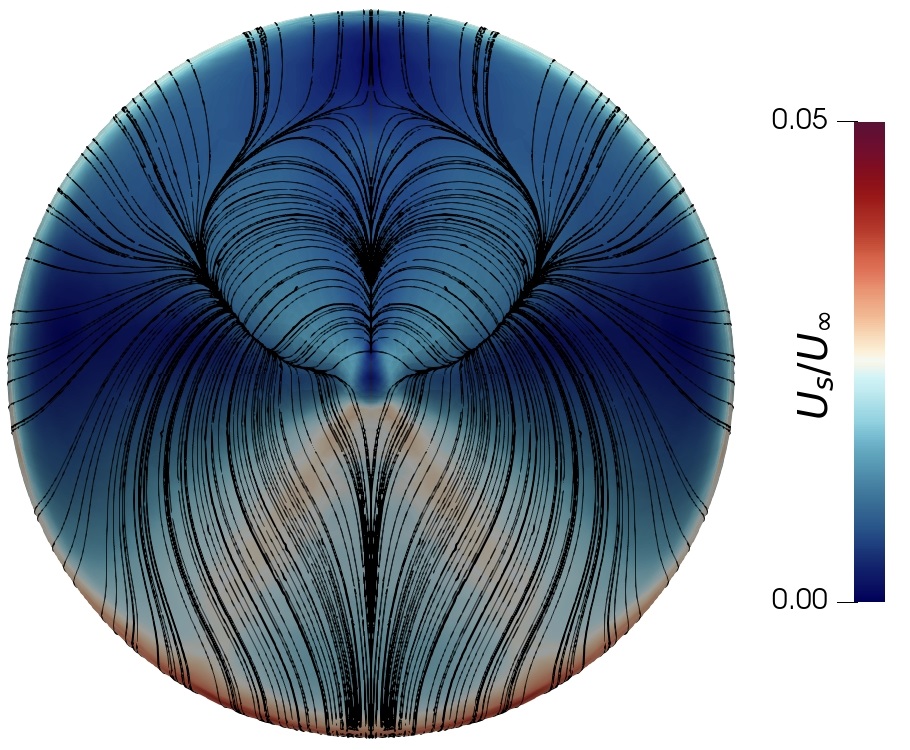}
    \caption{Surface oil flow streamlines overlaid on contours of surface slip velocity magnitude for the high-altitude flow conditions.}
    \label{fig:oil}
\end{figure}

The behavior of the flow at the surface was then visualized through surface velocity streamlines (i.e., oil flow). The contours of surface slip velocity as well as the oil flow streamlines are shown in \cref{fig:oil}. Due to the local non-equilibrium nature of the flow, non-negligible levels of surface slip velocity were observed, with the highest magnitude seen in the shoulder region. In this region, the slip velocity was on the order of $5\%$ of the freestream velocity. Another region of high slip velocity was observed on the windward side of the afterbody, although the magnitude was not to the extent of the slip velocity at the shoulder. 

The surface streamlines show a pronounced separation region on the leeward side of the afterbody. The separation region was largely convex and circular with a size on the order of half of the diameter of the capsule. This surface flow behavior was notably different than the predictions of \citet{Wright2006}, although these discrepancies are expected due to the different governing equations and wall boundary conditions (e.g., the present work does not assume no slip walls) as well as the differences in the physics being modeled. However, the location and size of the separation region was somewhat similar between the present work and the work of \citet{Wright2006}.

\subsubsection{Wake structure}
\begin{figure}[htbp!]
    \centering
    \includegraphics[width=0.7\textwidth]{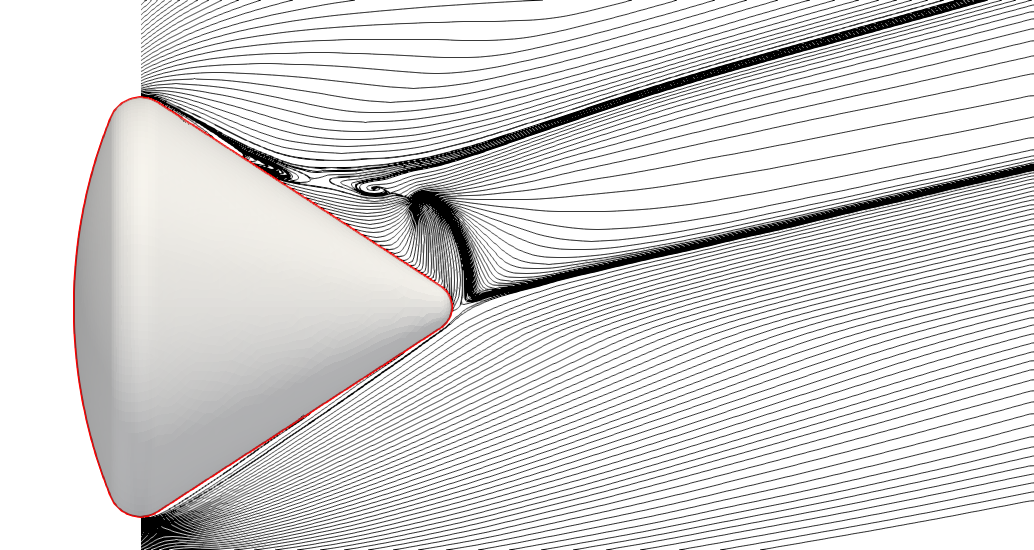}
    \caption{Surface streamlines of velocity in the wake region on the pitch plane ($z = 0$) for the high-altitude flow conditions. }
    \label{fig:stream}
\end{figure}

To observe the flow structure in the wake region, the velocity field on the pitch plane was first visualized through surface streamlines, shown in \cref{fig:stream}. The recompression shocks are evident in the streamlines, with two distinct regions of high velocity gradients aft of the capsule surface. Furthermore, a complex wake structure can be observed starting with the separation of the flow at the shoulder. This then develops into a small recirculation region on the leeward side of the capsule afterbody. The interaction between this recirculating region and the recompression shocks causes a complex rollup in the wake with a distinct critical point. 

\begin{figure}[htbp!]
    \centering
    \includegraphics[width=0.5\textwidth]{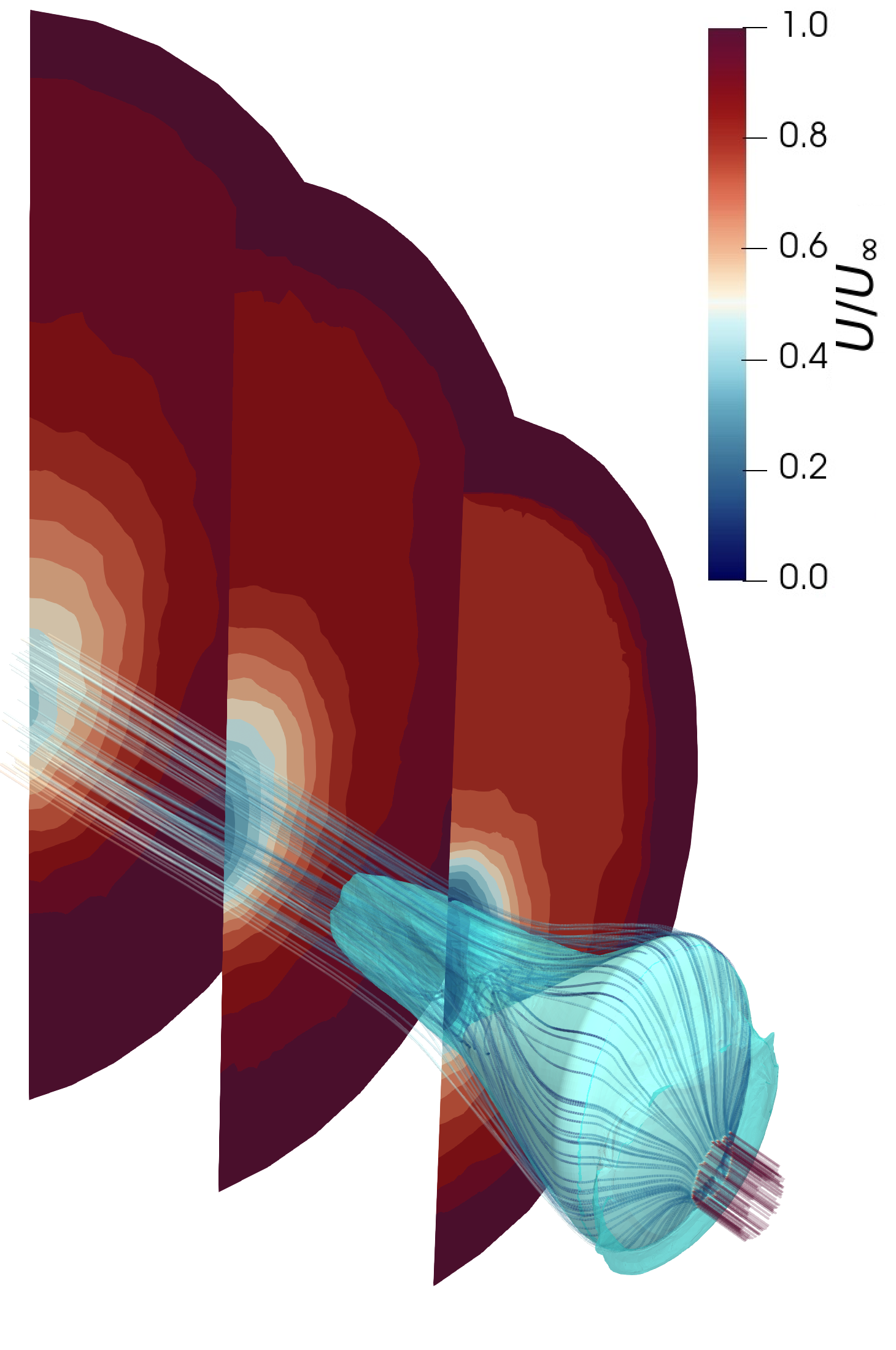}
    \caption{Three-dimensional visualization of velocity streamlines and cross-sections (placed 1, 2, and 3 capsule diameters downstream, respectively) colored by velocity magnitude for the high-altitude flow conditions. Sonic surface shown by light blue isosurface.}
    \label{fig:3d}
\end{figure}

The three-dimensional behavior of the flow was then visualized through flow streamlines and velocity cross-sections, shown in \cref{fig:3d}. The streamlines and velocity cross-sections evidently show the separation region aft of the capsule surface on the leeward side. Furthermore, the rollup of the recirculation region is also shown clearly in the streamlines. The sonic surface, the isosurface where $M = 1$, is shown in blue. This sonic surface encapsulates a majority of the wake region and shows similarity to the results of \citet{Wright2006}. 

\subsubsection{Shock structure \& non-equilibrium flow behavior}
\begin{figure}[htbp!]
    \centering
    \subfloat[Schlieren]{\includegraphics[width=0.5\textwidth]{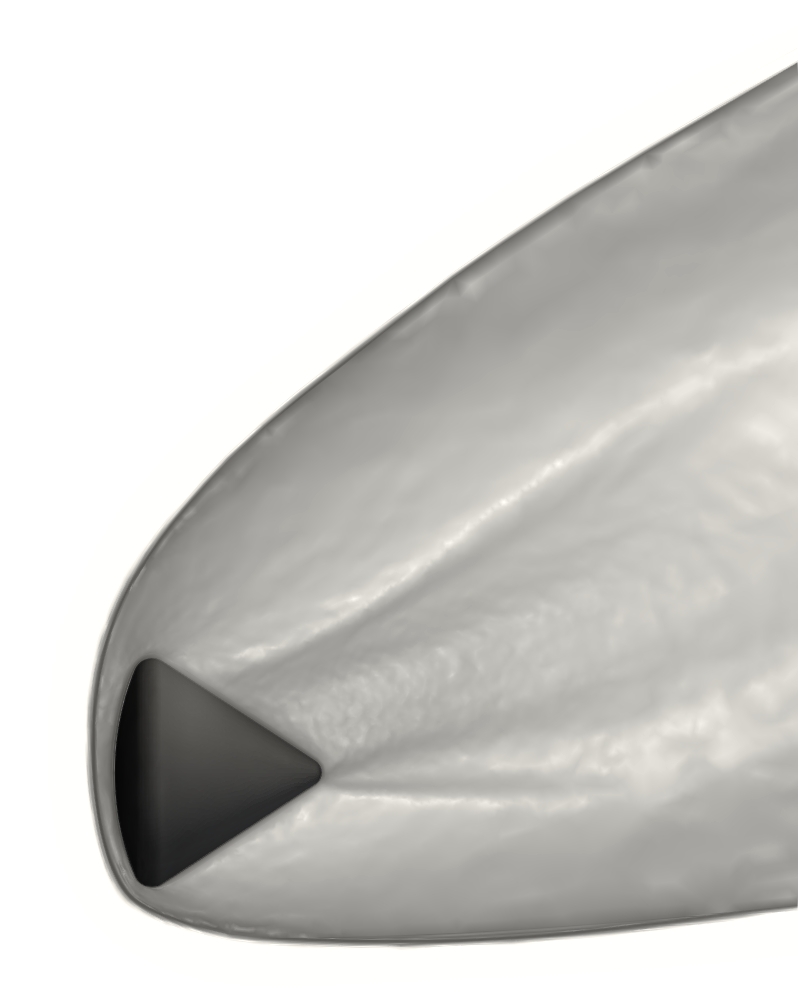}}
    \subfloat[Local Knudsen number]{\includegraphics[width=0.5\textwidth]{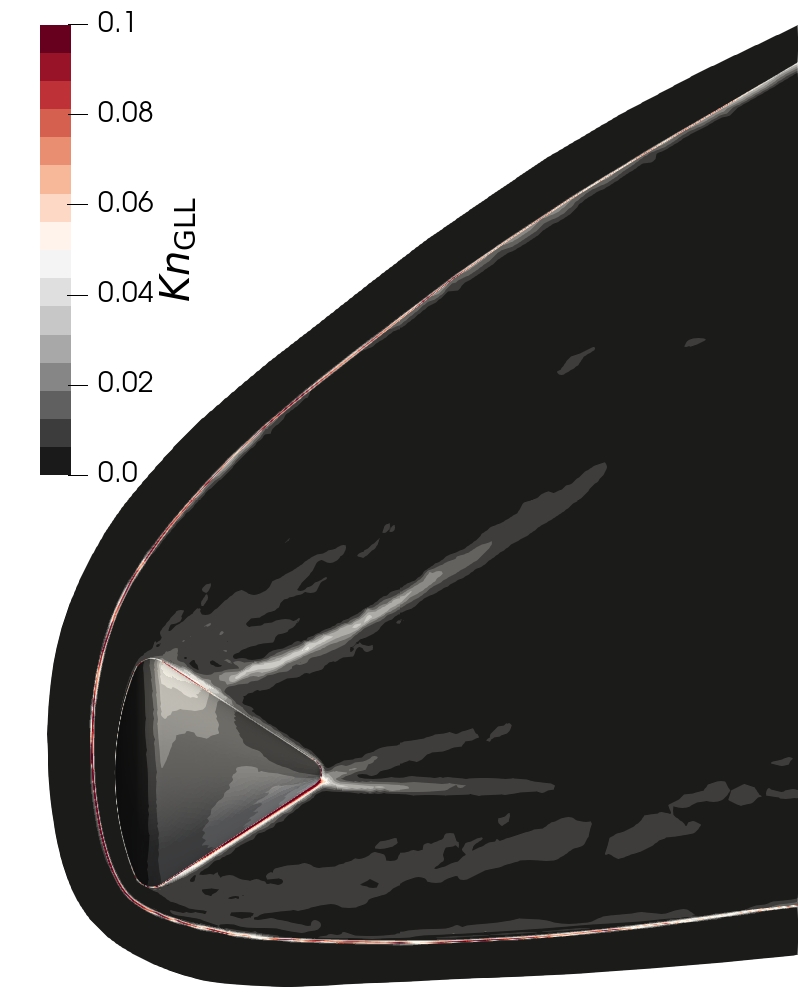}}
    \caption{Schlieren-type visualization of the density gradient norm (left) and contours of the density gradient-length local Knudsen number (right) on the pitch plane ($z = 0$) and the capsule surface for the high-altitude flow conditions. }
    \label{fig:schlieren}
\end{figure}

Finally, the shock structure and effects of flow non-equilibrium were analyzed. To visualize shocks in the flow, a Schlieren-type representation of the density gradient norm is shown in \cref{fig:schlieren}, visaulized on the pitch plane and capsule surface. The high density gradients in the bow shock are clearly visible as well as in the recompression region. Furthermore, it can be seen that even with a high-order scheme on an unstructured mesh, the numerical solution is still very well behaved around strong discontinuities, showcasing the ability of the approach in directly resolving shock structures without shock capturing schemes.

\begin{figure}[tbhp]
    \subfloat[Density]{\adjustbox{width=0.33\linewidth, valign=b}{\begin{tikzpicture}[spy using outlines={rectangle, height=3cm,width=2.3cm, magnification=3, connect spies}]
	\begin{axis}[name=plot1,
		axis line style={latex-latex},
	    axis x line=left,
        axis y line=left,
		xlabel={$x/D$},
    	xmin=-.15, xmax=-.05,
    	xtick={-0.05, -0.1, -0.15},
    	ylabel={$\rho/\rho_{\infty}$},
    	ymin=0,ymax=10,
        clip mode=individual,
        x tick label style={/pgf/number format/.cd, fixed, fixed zerofill, precision=2, /tikz/.cd},
    	legend style={at={(0.97, 0.1)},anchor=south east,font=\small},
    	legend cell align={left},
    	style={font=\normalsize}]
        
        \addplot[color=black, 
                style={very thick}]
        table[x=x, y=d, col sep=comma]{./figs/tikz/shock_profile_post.csv};

	\end{axis}
\end{tikzpicture}}}
    \subfloat[Velocity]{\adjustbox{width=0.33\linewidth, valign=b}{\begin{tikzpicture}[spy using outlines={rectangle, height=3cm,width=2.3cm, magnification=3, connect spies}]
	\begin{axis}[name=plot1,
		axis line style={latex-latex},
	    axis x line=left,
        axis y line=left,
		xlabel={$x/D$},
    	xmin=-.15, xmax=-.05,
    	xtick={-0.05, -0.1, -0.15},
    	ylabel={$\|U\|/\|U\|_{\infty}$},
    	ymin=0,ymax=1.1,
        clip mode=individual,
        x tick label style={/pgf/number format/.cd, fixed, fixed zerofill, precision=2, /tikz/.cd},
    	legend style={at={(0.97, 0.1)},anchor=south east,font=\small},
    	legend cell align={left},
    	style={font=\normalsize}]
        
        \addplot[color=black, 
                style={very thick}]
        table[x=x, y=v, col sep=comma]{./figs/tikz/shock_profile_post.csv};

	\end{axis}
\end{tikzpicture}}}
    \subfloat[Pressure]{\adjustbox{width=0.33\linewidth, valign=b}{\begin{tikzpicture}[spy using outlines={rectangle, height=3cm,width=2.3cm, magnification=3, connect spies}]
	\begin{axis}[name=plot1,
		axis line style={latex-latex},
	    axis x line=left,
        axis y line=left,
		xlabel={$x/D$},
    	xmin=-.15, xmax=-.05,
    	xtick={-0.05, -0.1, -0.15},
    	ylabel={$P/P_{\infty}$},
    	ymin=0,ymax=800,
        clip mode=individual,
        x tick label style={/pgf/number format/.cd, fixed, fixed zerofill, precision=2, /tikz/.cd},
    	legend style={at={(0.97, 0.1)},anchor=south east,font=\small},
    	legend cell align={left},
    	style={font=\normalsize}]
        
        \addplot[color=black, 
                style={very thick}]
        table[x=x, y=p, col sep=comma]{./figs/tikz/shock_profile_post.csv};

	\end{axis}
\end{tikzpicture}}}
    \newline
    \caption{\label{fig:shock_profiles} Profiles of normalized density (left), velocity magnitude (middle), and pressure (right) across the bow shock in front of the capsule nose ($y = 0$, $z = 0$) for the high-altitude flow conditions. 
    }
\end{figure}
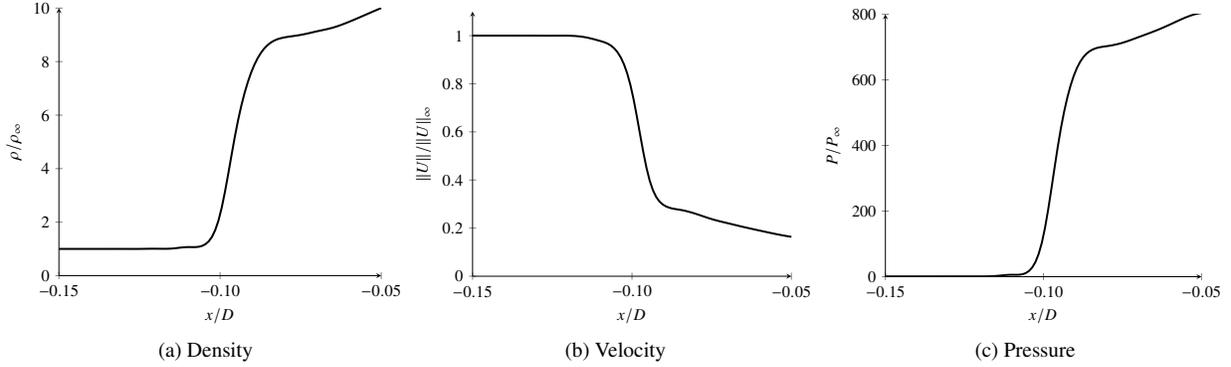

The behavior of the flow across the bow shock is shown more explicitly in the flow profiles in front of the capsule nose in \cref{fig:shock_profiles}. Prior to the compression in the flow near the capsule surface, the behavior of the shock is similar to a static one-dimensional normal shock. It can be seen that the large jumps in density and pressure are well-resolved by the numerical scheme without any spurious oscillations. Furthermore, from the validation work performed in \citet{Dzanic2023}, it is expected that the scheme predicts the physically correct shock thickness and behavior. As can be seen in \cref{fig:shock_profiles}, the shock thickness is of $\mathcal O(10^{-2}D)$, which is consistent with an upstream mean free path of $\mathcal O(10^{-3}D)$ and falls well within the resolving capability of the given mesh and approximation order. 

To quantify the degree of non-equilibrium in the flow, the density gradient-length local Knudsen number was computed, given as 
\begin{equation}
    Kn_{\text{GLL}} = \lambda \frac{\|\nabla \rho \|_2}{\rho},
\end{equation}
where $\lambda$ is the local mean free path defined as 
\begin{equation}
    \lambda = \frac{\mu (\theta)}{\rho}\sqrt{\frac{\pi}{2\theta}.}
\end{equation}
From \citet{Boyd1995}, it is expected that continuum breakdown begins when $Kn_{\text{GLL}} > 0.05$. The contours of the local Knudsen number are also shown on the pitch plane and capsule surface in \cref{fig:schlieren}. It can be seen that this breakdown condition is exceeded in a large portion of the flow. In particular, the bow shock shows a very large local Knudsen number, with a maximum of nearly $0.5$. Furthermore, the capsule surface on the leeward side near the shoulder also shows flow regions which exceed this condition as well as the recompression shock region. These results are in line with the observations of \citet{Wright2006} which may explain the discrepancies in the results obtained by \citet{Wright2006} using the continuum fluid dynamics equations in comparison to the experimental data.

\subsection{Low-altitude conditions}
\subsubsection{Macroscopic flow cross-sections}
\begin{figure}[htbp!]
    \centering
    \subfloat[Density]{\includegraphics[width=0.48\textwidth]{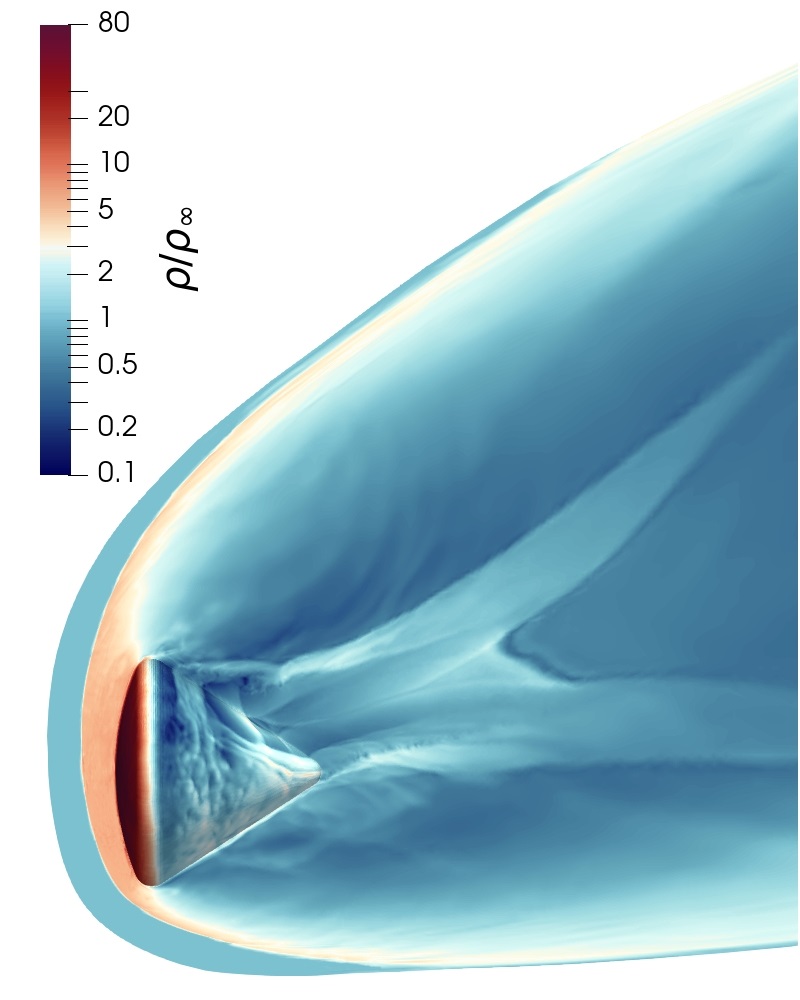}}
    \subfloat[Pressure]{\includegraphics[width=0.48\textwidth]{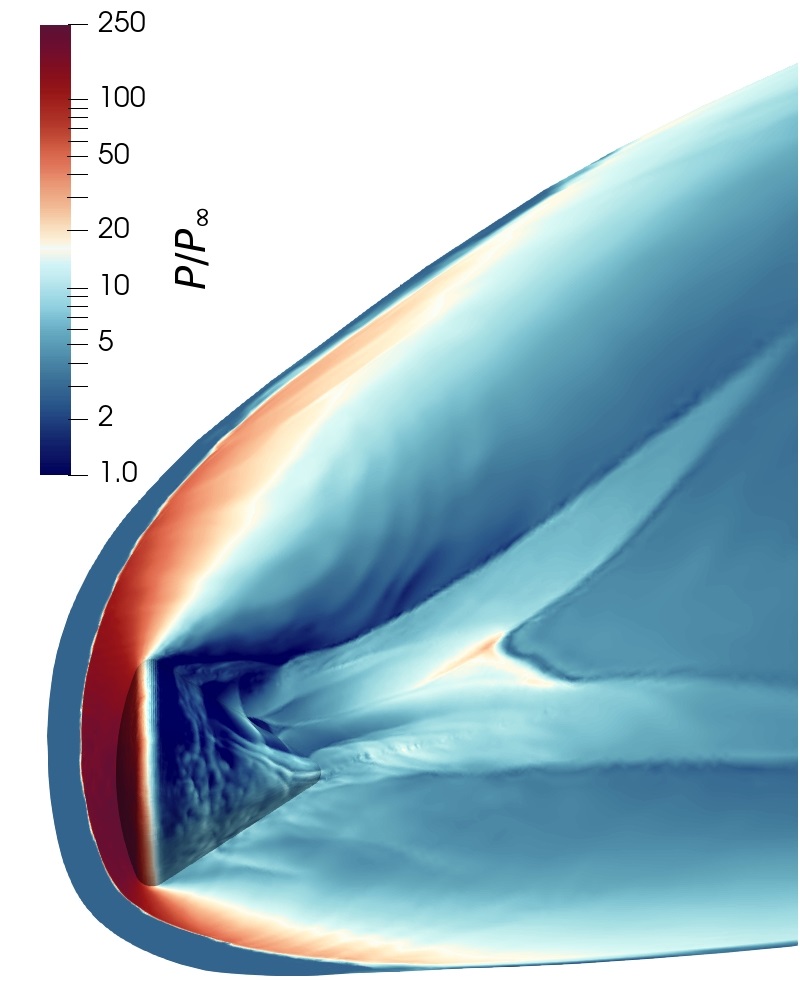}}
    \newline
    \subfloat[Velocity]{\includegraphics[width=0.48\textwidth]{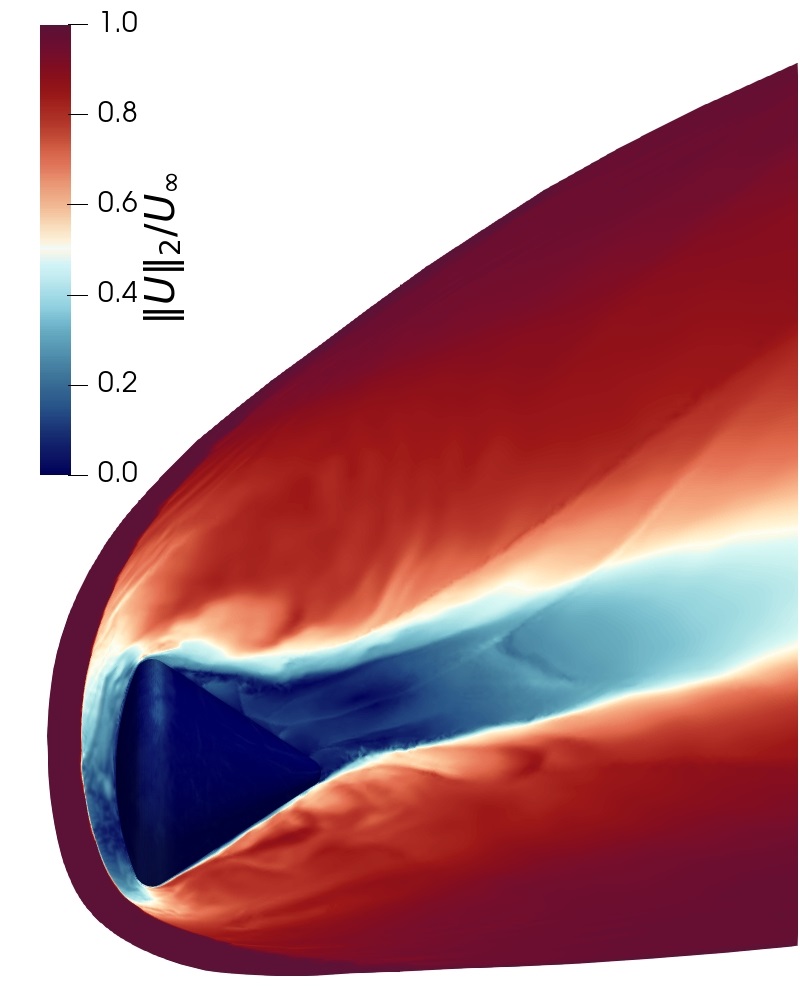}}
    \subfloat[Temperature]{\includegraphics[width=0.48\textwidth]{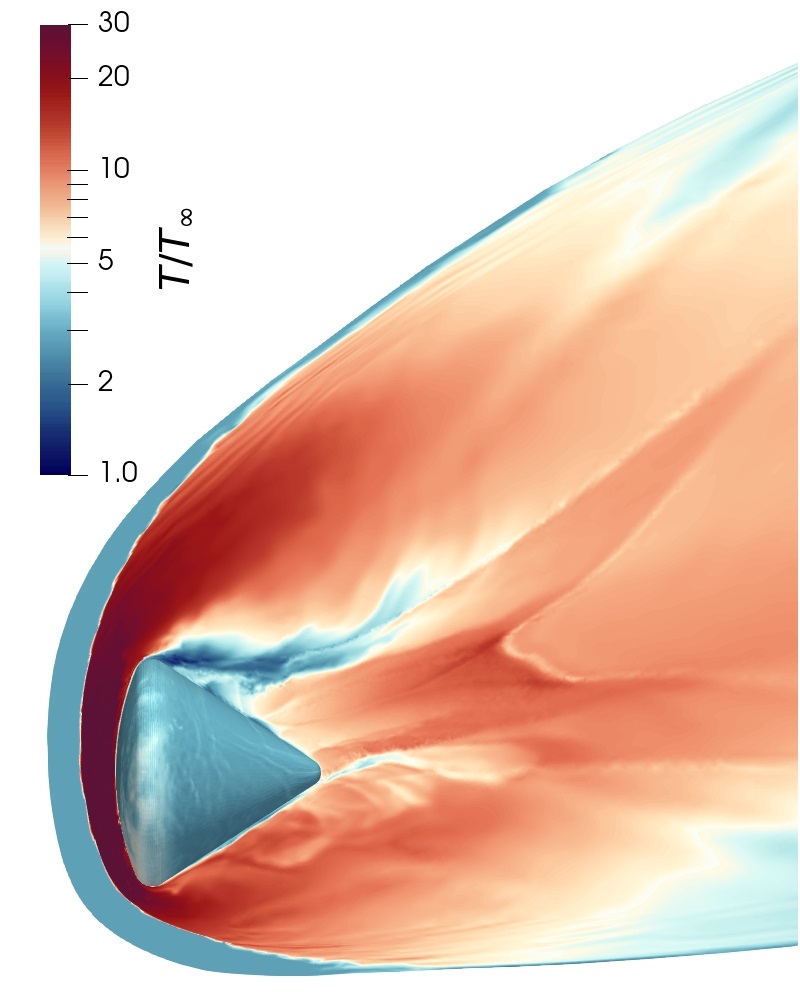}}
    \newline
    \caption{Instantaneous contours of macroscopic normalized density (top left), pressure (top right), velocity magnitude (bottom left), and temperature (bottom right) on the pitch plane ($z = 0$) and the capsule surface for the low-altitude flow conditions. }
    \label{fig:sol2}
\end{figure}

The results of the unsteady, low-altitude, low Mach simulation were then analyzed with respect to the \emph{instantaneous} macroscopic flow state on the symmetry plane cross-section ($z = 0$). The contours of instantaneous density, pressure, velocity magnitude, and temperature are shown in \cref{fig:sol2} normalized by the freestream values. It can be seen that the structures in the flow are significantly more complex than the high-altitude case, largely as a result of the higher Reynolds number driving an unsteady turbulent wake flow. A similar separation region could be seen aft of the capsule, but the recompression shocks were noticeably sharper and showed more complex features which may also be attributed to the higher Reynolds number and, by extension, lower Knudsen number in the flow. The contours of density and pressure show a transitional boundary layer aft of the shoulder which develop into unsteady flow features in the wake. Similarly to the high-altitude case, the bow shock was evident with large jumps in the macroscopic flow variables across the shock. Furthermore, even when using a mesh not aligned with the shock, the scheme was still able to robustly and accurately resolve the strong shock without any explicit shock capturing approach. This showcases the robustness of the approach, particularly for complex flows and flows which may span many flow regimes as the method is not noticeably sensitive to mesh quality/design which is typically a drawback for high-order numerical schemes.

\subsubsection{Wake structure}
\begin{figure}[htbp!]
    \centering
    \includegraphics[width=0.7\textwidth]{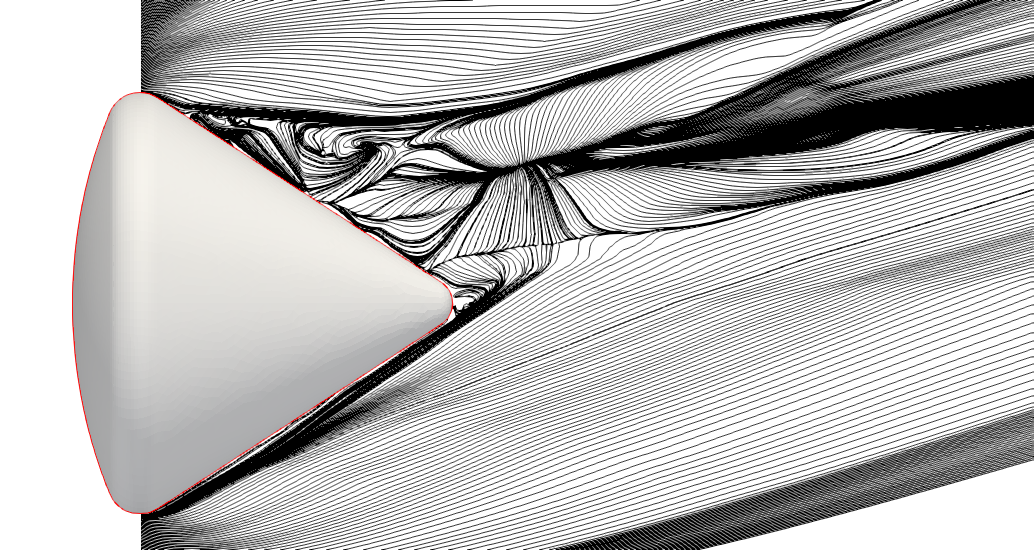}
    \caption{Instantaneous surface streamlines of velocity in the wake region on the pitch plane ($z = 0$) for the low-altitude flow conditions. }
    \label{fig:stream2}
\end{figure}

The flow structure in the wake region was similarly analyzed by visualizing the \emph{instantaneous} surface streamlines on the pitch plane, shown in \cref{fig:stream2}. It can be seen that the unsteady turbulent wake flow results in significantly more complex behavior in the flow streamlines. A noticeably larger separation region on the leeward side was observed than the high-altitude case, likely as a result of the lower wall slip velocity in the near-continuum regime, but the flow separated at a similar location aft of the shoulder. The streamlines show very chaotic flow in this wake, with large flow gradients around the recompression shocks and the distinct rollup of vortical flow features. On the windward side of the capsule, the flow streamlines were primarily smooth, but the near-wall region showed signs of transition to turbulence. These results highlight the ability of the approach in resolving complex unsteady flow features which high-order numerical schemes are particularly well-suited for. 

\section{Conclusion}\label{sec:conclusion}
We present the results of high-order numerical experiments of hypersonic re-entry vehicles computed by performing three-dimensional molecular gas dynamics simulations through directly solving the six-dimensional Boltzmann --BGK equation. The underlying numerical approach combines a highly-efficient high-order unstructured discontinuous spectral element method for the spatial domain with a discretely-conservative discrete velocity model for the velocity domain. This approach was used to simulate the AS-202 Apollo capsule at realistic re-entry conditions in both the high-altitude regime at a Mach number of 22.7 and a Reynolds number of 43,000 and the low-altitude regime at a Mach number of 8 and a Reynolds number of 550,000. These results demonstrate a proof of feasibility of the approach on modern computing architectures, showcasing the ability to directly resolve strong shocks, non-equilibrium flow phenomena, and unsteady turbulent flow features. Future work will focus on the implementation of various models to better approximate the underlying physics in high-speed re-entry flows. 


\section*{Acknowledgments}

This work was partially performed under the auspices of the U.S. Department of Energy by Lawrence Livermore National Laboratory under contract DE--AC52--07NA27344. Release number LLNL--CONF--858000. The authors would like to acknowledge the computational resources provided by the Princeton Institute for Computational Science and Engineering and by Professor Freddie Witherden at Texas A\&M University.

\bibliography{sample}

\begin{thebibliography}{24}
\newcommand{\enquote}[1]{``#1''}
\providecommand{\natexlab}[1]{#1}
\providecommand{\url}[1]{\texttt{#1}}
\providecommand{\urlprefix}{URL }
\expandafter\ifx\csname urlstyle\endcsname\relax
  \providecommand{\doi}[1]{\discretionary{}{}{}https://doi.org/#1}\else
  \providecommand{\doi}[1]{\discretionary{}{}{}\urlstyle{rm}\url{https://doi.org/#1}}\fi

\bibitem[{{Schneider}(2006)}]{Schneider2006}
{Schneider}, S.~P., \enquote{{Laminar-Turbulent Transition on Reentry Capsules and Planetary Probes},} \emph{Journal of Spacecraft and Rockets}, Vol.~43, No.~6, 2006, pp. 1153--1173.
\newblock \doi{10.2514/1.22594}.

\bibitem[{Dzanic et~al.(2023{\natexlab{a}})Dzanic, Witherden, and Martinelli}]{Dzanic2023}
Dzanic, T., Witherden, F., and Martinelli, L., \enquote{A positivity-preserving and conservative high-order flux reconstruction method for the polyatomic {B}oltzmann{\textendash}{BGK} equation,} \emph{Journal of Computational Physics}, Vol. 486, 2023{\natexlab{a}}, p. 112146.
\newblock \doi{10.1016/j.jcp.2023.112146}.

\bibitem[{Dzanic et~al.(2023{\natexlab{b}})Dzanic, Witherden, and Martinelli}]{Dzanic2023b}
Dzanic, T., Witherden, F.~D., and Martinelli, L., \enquote{Validation of wall boundary conditions for simulating complex fluid flows via the {B}oltzmann--{BGK} equation: Momentum transport and skin friction,} \emph{arXiv}, 2023{\natexlab{b}}.

\bibitem[{Wright et~al.(2006)Wright, Prabhu, and Martinez}]{Wright2006}
Wright, M.~J., Prabhu, D.~K., and Martinez, E.~R., \enquote{Analysis of {A}pollo Command Module Afterbody Heating Part I: {AS}-202,} \emph{Journal of Thermophysics and Heat Transfer}, Vol.~20, No.~1, 2006, pp. 16--30.
\newblock \doi{10.2514/1.15873}.

\bibitem[{Yuan and Zhong(2020)}]{Yuan2020}
Yuan, R., and Zhong, C., \enquote{A conservative implicit scheme for steady state solutions of diatomic gas flow in all flow regimes,} \emph{Computer Physics Communications}, Vol. 247, 2020, p. 106972.
\newblock \doi{10.1016/j.cpc.2019.106972}.

\bibitem[{Zhang et~al.(2023)Zhang, Zeng, Yuan, Liu, Li, and Wu}]{Zhang2023}
Zhang, Y., Zeng, J., Yuan, R., Liu, W., Li, Q., and Wu, L., \enquote{Efficient parallel solver for high-speed rarefied gas flow using GSIS,} , 2023.
\newblock \doi{10.48550/ARXIV.2310.18916}.

\bibitem[{Chen et~al.(2020)Chen, Zhu, and Xu}]{Chen2020}
Chen, Y., Zhu, Y., and Xu, K., \enquote{A three-dimensional unified gas-kinetic wave-particle solver for flow computation in all regimes,} \emph{Physics of Fluids}, Vol.~32, No.~9, 2020.
\newblock \doi{10.1063/5.0021199}.

\bibitem[{Cercignani(1988)}]{Cercignani1988}
Cercignani, C., \emph{The Boltzmann Equation and Its Applications}, Springer New York, 1988.
\newblock \doi{10.1007/978-1-4612-1039-9}.

\bibitem[{Bhatnagar et~al.(1954)Bhatnagar, Gross, and Krook}]{Bhatnagar1954}
Bhatnagar, P.~L., Gross, E.~P., and Krook, M., \enquote{A Model for Collision Processes in Gases. {I}. {S}mall Amplitude Processes in Charged and Neutral One-Component Systems,} \emph{Physical Review}, Vol.~94, No.~3, 1954, pp. 511--525.
\newblock \doi{10.1103/physrev.94.511}.

\bibitem[{Holway(1966)}]{Holway1966}
Holway, L.~H., \enquote{New Statistical Models for Kinetic Theory: Methods of Construction,} \emph{Physics of Fluids}, Vol.~9, No.~9, 1966, p. 1658.
\newblock \doi{10.1063/1.1761920}.

\bibitem[{Baranger et~al.(2020)Baranger, Dauvois, Marois, Math{\'{e}}, Mathiaud, and Mieussens}]{Baranger2020}
Baranger, C., Dauvois, Y., Marois, G., Math{\'{e}}, J., Mathiaud, J., and Mieussens, L., \enquote{A {BGK} model for high temperature rarefied gas flows,} \emph{European Journal of Mechanics - B/Fluids}, Vol.~80, 2020, pp. 1--12.
\newblock \doi{10.1016/j.euromechflu.2019.11.006}.

\bibitem[{Sutherland(1893)}]{Sutherland1893}
Sutherland, W., \enquote{{LII}. The viscosity of gases and molecular force,} \emph{The London, Edinburgh, and Dublin Philosophical Magazine and Journal of Science}, Vol.~36, No. 223, 1893, p. 507–531.
\newblock \doi{10.1080/14786449308620508}.

\bibitem[{Huynh(2007)}]{Huynh2007}
Huynh, H.~T., \enquote{A Flux Reconstruction Approach to High-Order Schemes Including Discontinuous {G}alerkin Methods,} \emph{18th {AIAA} Computational Fluid Dynamics Conference}, American Institute of Aeronautics and Astronautics, 2007.
\newblock \doi{10.2514/6.2007-4079}.

\bibitem[{Hesthaven and Warburton(2008)}]{Hesthaven2008DG}
Hesthaven, J.~S., and Warburton, T., \emph{Nodal Discontinuous {G}alerkin Methods}, Springer New York, 2008.
\newblock \doi{10.1007/978-0-387-72067-8}.

\bibitem[{Cai and Torrilhon(2019)}]{Cai2019}
Cai, Z., and Torrilhon, M., \enquote{On the {H}olway-{W}eiss debate: Convergence of the {G}rad-moment-expansion in kinetic gas theory,} \emph{Physics of Fluids}, Vol.~31, No.~12, 2019.
\newblock \doi{10.1063/1.5127114}.

\bibitem[{Loek Van~Heyningen(2021)}]{Heyningen2021}
Loek Van~Heyningen, R., \enquote{Discontinuous {G}alerkin solutions of the {B}oltzmann equation: spectral collocation and moment methods,} Master's thesis, Massachusetts Institute of Technologyy, 2021.

\bibitem[{Mieussens(2000{\natexlab{a}})}]{Mieussens2000}
Mieussens, L., \enquote{Discrete velocity model and implicit scheme for the {BGK} equation of rarefied gas dynamic,} \emph{Mathematical Models and Methods in Applied Sciences}, Vol.~10, No.~08, 2000{\natexlab{a}}, pp. 1121--1149.
\newblock \doi{10.1142/s0218202500000562}.

\bibitem[{Trefethen and Weideman(2014)}]{Trefethen2014}
Trefethen, L.~N., and Weideman, J. A.~C., \enquote{The Exponentially Convergent Trapezoidal Rule,} \emph{{SIAM} Review}, Vol.~56, No.~3, 2014, pp. 385--458.
\newblock \doi{10.1137/130932132}.

\bibitem[{Mieussens(2000{\natexlab{b}})}]{Mieussens2000JCP}
Mieussens, L., \enquote{Discrete-Velocity Models and Numerical Schemes for the {B}oltzmann-{BGK} Equation in Plane and Axisymmetric Geometries,} \emph{Journal of Computational Physics}, Vol. 162, No.~2, 2000{\natexlab{b}}, pp. 429--466.
\newblock \doi{10.1006/jcph.2000.6548}.

\bibitem[{Hillje(1967)}]{Hillje1967}
Hillje, E.~R., \enquote{Entry flight aerodynamics from Apollo mission AS-202,} Tech. Rep. 19670027745, Manned Spacecraft Center, NASA, Houston, Texas, October 1967.

\bibitem[{Harten et~al.(1983)Harten, Lax, and Leer}]{Harten1983}
Harten, A., Lax, P.~D., and Leer, B.~v., \enquote{On Upstream Differencing and {G}odunov-Type Schemes for Hyperbolic Conservation Laws,} \emph{SIAM Review}, Vol.~25, No.~1, 1983, p. 35–61.
\newblock \doi{10.1137/1025002}.

\bibitem[{Dzanic and Witherden(2022)}]{Dzanic2022}
Dzanic, T., and Witherden, F., \enquote{Positivity-preserving entropy-based adaptive filtering for discontinuous spectral element methods,} \emph{Journal of Computational Physics}, Vol. 468, 2022, p. 111501.
\newblock \doi{10.1016/j.jcp.2022.111501}.

\bibitem[{Witherden et~al.(2014)Witherden, Farrington, and Vincent}]{Witherden2014}
Witherden, F., Farrington, A., and Vincent, P., \enquote{{PyFR}: An open source framework for solving advection{\textendash}diffusion type problems on streaming architectures using the flux reconstruction approach,} \emph{Computer Physics Communications}, Vol. 185, No.~11, 2014, pp. 3028--3040.
\newblock \doi{10.1016/j.cpc.2014.07.011}.

\bibitem[{Boyd et~al.(1995)Boyd, Chen, and Candler}]{Boyd1995}
Boyd, I.~D., Chen, G., and Candler, G.~V., \enquote{Predicting failure of the continuum fluid equations in transitional hypersonic flows,} \emph{Physics of Fluids}, Vol.~7, No.~1, 1995, p. 210–219.
\newblock \doi{10.1063/1.868720}.

\end{thebibliography}

\end{document}